\documentclass{elsart}

\usepackage{psfig}

\usepackage{natbib}

\begin{document}

\runauthor{Gaskell}

% -------------------------------------------------------------------------

\begin{frontmatter}

\title{A Look at What Is (and Isn't) Known About Quasar Broad Line Regions and How
Narrow-Line Seyfert 1 Galaxies Fit In}

\author[UNL]{C.\,Martin Gaskell}
\address[UNL]{Department of Physics \& Astronomy,
University of Nebraska, Lincoln, NE 68588-0111, USA}

\begin{abstract}
The evidence is reviewed that the Broad Line Region (BLR) probably
has two distinct components located at about the same distance
from the central black hole. One component, BLR~II, is
optically-thick, low-ionization emission at least some of which
arises from a disc and the other, BLR~I, is probably
optically-thin emission from a more spherically symmetric halo or
atmosphere. The high Fe~II/H$\beta$ ratios seen in Narrow-Line
Seyfert 1 galaxies (NLS1s) are not due to strong Fe~II emission,
as is commonly thought, but to unusually weak Balmer emission,
probably caused by higher densities. NLS1s probably differ from
non-NLS1s because of the higher density of gas near the black
hole.  This produces a higher accretion rate, a denser BLR, and a
view of the central regions that is more face-on.

\end{abstract}


\end{frontmatter}

% -------------------------------------------------------------------------

\section{Introduction}

I was asked by the organizers if I would give ``some kind of
review of the properties of the BLR.'' Why, at a workshop on NLS1s,
should we consider BLRs in non-NLS1s?  For me, and perhaps most
people who attended this workshop, the interest of NLS1s has been
that these extrema in the distributions of AGN properties might
tell us more about how {\it all} AGNs work.  This is, however, a
two-way street: if NLS1 are ``just'' extrema of a continuum of BLR
properties, then NLS1 BLRs are not {\it fundamentally} different
from the BLRs of other AGNs. Therefore, a model of NLS1s {\it must
have the BLRs be consistent with those of non-NLS1s}. I present
here an overview of some general BLR results with which
NLS1 models need to be consistent.

In BLR research the big question is ``what role does the BLR play
in the quasar phenomenon?''  Some very basic specific questions
include:
\begin{itemize}
\item Is the BLR one thing or more than one thing?
\item Where is the BLR located? Or where are the components of the BLR
located? ({\it e.g.}, what is the distribution in $R, \Theta,
\Phi$?)
\item Where is the BLR coming from or going to?
\end{itemize}

A survey of participants in the 1998 Nebraska BLR conference
revealed a complete lack of consensus on the answers to these
basic questions (Gaskell 1999).

\section{Basic BLR Parameters}

Observationally an AGN presents the astronomer with the four
Stokes parameters as a function of projected velocity ($v_{\rm proj}$)
and time ($t$).  If we ignore polarization and subtract out
various problems such as the continuum, absorption lines, and
blended emission lines, then, for each line, $i$, we get
$L_{i}(v_{\rm proj},t)$.  If we ignore $t$ we get $L_{i}(v_{\rm proj})$,
the line profiles.  If we ignore $v$ we get $L_i$, the integrated
intensities of the lines.

The traditional approach (going back to the late 1960s) has been
to try to explain $L_i$ with a ``typical'' photoionized cloud.
Obviously this assumes that the profiles are the same.  We will
see below that they are not, but this traditional approach has
none the less been quite successful in explaining the stronger
high-ionization lines and has produced a number of important
results.
\begin{itemize}
\item The broad lines are produced mainly by {\it
photoionization}. We can say this with confidence because {\it the
lines vary with the continuum}, and the lines can vary in
intensity by an order of magnitude.  Not only do the
high-ionization lines vary with the continuum, but the low
ionization lines (such as Mg~II and Fe~II) do too. Other energy
input sources might be present, but the dominant energy input is
from photoionization.
\item The mean physical conditions in the BLR do not vary a lot
from object to object.
\item The covering factor, $\Omega \sim 10\%$.
\item The density is $\sim 10^{10}$--$10^{12}$ (or $10^{13}$)
cm$^{-3}$.
\item The mass of gas seen in the BLR is of the order of a solar mass.
\item If the BLR emission arises in free clouds with
dimensions comparable the the Str\"omgren length, the number of
clouds is enormous ($\sim 10^{16}$)
\item Abundances are $\sim$ solar to a few times solar ({\it e.g.}, Gaskell,
Wampler, \& Shields 1981; Hamann \& Ferland 1993) which is
consistent with what one expects in the centers of galaxies.
\end{itemize}

\section{Optical Fe~II Strengths in NLS1s}

One of the major challenges for photo-ionization models has always
been to explain the great strengths of low-ionization lines. In
particular, explaining the origin of Fe~II emission is a
long-standing problem.  This is reviewed by Suzy Collin-Souffrin
elsewhere in this volume so I will not discuss it here, but I will
point out one result (Gaskell 1985) which is not widely
appreciated in discussion of the BLR of NLS1s.

One supposed characteristic of NLS1s is the great strength of the
optical Fe~II emission ({\it e.g.}, Boller, Brandt, \& Fink 1996).
Wills (1982) pointed out for a heterogenous collection of AGNs
that Fe~II/H$\beta$ seemed to be roughly inversely proportional to
the FWHM. As is well known, optical Fe~II is hard to measure.  I
therefore looked at H$\beta$ and optical Fe~II emission in the
Lick Observatory sample of Seyfert 1 spectra of Osterbrock and his
graduate students -- a sample with very high signal-to-noise ratio
spectra where the lines had been carefully de-blended.  From this
homogeneous sample I was able to confirm the FWHM {\it vs.}
Fe~II/H$\beta$ correlation of Wills (1982). However, the interesting
thing is that while there is considerable object-to-object
variation in the equivalent width of the optical Fe~II (something
that needs to be explained), {\it there is no correlation of the
equivalent width of the Fe~II emission with FWHM} (see figure 2 of
Gaskell 1985). Instead, the FWHM {\it vs.} Fe~II/H$\beta$ 
correlation arises {\it because H$\beta$ gets weaker as the lines get
narrower}.  So rather than trying to explain a (non-existent)
anomalous strength of Fe~II in NLS1s, we should be trying to
explain the anomalous weakness of H$\beta$ in NLS1s. H$\beta$ is
probably being thermalized because of the high density and this
led to the prediction (Gaskell 1985) that UV spectra of
narrow-line objects would show an increase in the strength of
Si~III] relative to C~III]. This prediction has now been confirmed
(Kuraszkiewicz {\it et al.} 1998; Wills {\it et al.} 1999).

\section{The Need For (At Least) Two BLR Components}

To be valid the traditional ``typical photoionized cloud''
analysis requires the $L_i(v)$ to be the same.  If we look at the
first moments of the line profiles (the line centroids) we find
that the high-ionization lines are blueshifted relative to both
the low-ionization lines and the rest frame of the host galaxy
(Gaskell 1982).  If we look at the second moments of the line
profiles (the line widths) we find that FWHM $\sim$ ionization
potential (Shuder 1982). These differences and other
considerations have led some workers to argue that there are at
least two fundamentally different BLR components (Gaskell 1987;
Collin-Souffrin \& Lasota 1988). Using the terminology of 
Gaskell (1987) these are:

\begin{itemize}
\item {\bf BLR I}---a fairly traditional H~II region producing the
strong UV lines ({\it e.g.}, as in the models of Davidson 1973).
Collin-Souffrin \& Lasota (1988) refer to BLR~I as the ``HIL''
(high-ionization) BLR.
\item {\bf BLR II}---a large partially ionized zone (PIZ) which
produces strong Balmer emission, Mg~II, Fe~II, Ca~II, O~I, {\it
etc.}.  Collin-Souffrin \& Lasota (1988) refer to this as the ``LIL''
(low-ionization) BLR.
\end{itemize}

There are several reasons why I believe we need two (or more)
components:

1. {\it To Explain the Integrated Intensities}

Collin-Souffrin {\it et al.} (1979, 1980, 1981) have argued that
single photo-ionized clouds are unable to simultaneously explain
the integrated line intensities of both the high-ionization lines
and the hydrogen lines (especially Ly$\alpha$/H$\beta$).  This is true
{\it a fortiori} when one considers $L_i(v)$ (Snedden \& Gaskell
1999{\it ab}, 2000). By constructing a grid of models using Gary
Ferland's photoionization code CLOUDY (Ferland 2000) and looking
at the profiles of the strong BLR~I lines, we get a hydrogen
density $n_H \sim$ const. $(\sim 10^{11}$ cm$^{-3})$, independent
of the projected velocity, $v$, and an ionization parameter, $U_1
\sim$ const. $(\sim 10^{-1.5})$, again independent of $v$. We
found that the BLR~I lines alone can be explained by either
optically thick or optically thin models.

When we take the physical conditions we deduce from our analysis
of the BLR~I lines and try to predict the strengths of the BLR~II
lines we find that optically thick photoionization models can only
explain the hydrogen-line ratios at low $v$; the {\it wings} of
the Balmer lines are seriously overpredicted (see Figure 2 in
Gaskell \& Snedden 1999$a$).  This means that BLR~I {\it is mostly
optically thin}.  This has been already suggested by Ferland,
Korista, \& Peterson (1990) from emission-line variability
considerations.  When we try to use our grids of CLOUDY models to
deduce conditions from the BLR~II lines alone we find we need very
optically thick clouds with $n_H \sim 10^{13}$ cm$^{-3}$ and a
very low ionization parameter, $U_1 \sim 10^{-3} - 10^{-4}$ to
satisfy the constraints (Snedden \& Gaskell 2000).

2. {\it To Explain the Different Line Profiles}

(a) As already noted above, BLR~I lines are broader than BLR~II
lines.

(b) Not only are the line widths different, but Mathews \& Wampler
(1985) found the C~IV (BLR~I) and Mg~II (BLR~II) FWHMs in general
to be uncorrelated for both radio-loud and radio-quiet AGNs.
Further analysis (Gaskell \& Mariupolskaya 2000), while confirming
the independence of the C~IV and Mg~II FWHMs, shows that the
relationship between BLR~I and BLR~II FWHMs is complicated (for
example, Corbin 1993 found the FWHMs of C~IV and H$\beta$ to be
quite well correlated).

3. {\it Because BLR I is Blueshifted}

The standard explanation of the blueshifting (Gaskell 1982) is
that BLR~I is (at least approximately) spherical and radially
outflowing but something (the accretion disc or inner torus) is
blocking our view of the far side.  There are, however, problems
with this and other explanations have been offered (see below).

4. {\it Because BLR I and BLR II Emission Come From Different
Radii}

One of the first results of what is often called ``reverberation
mapping'' of BLRs (using light echoes to probe the structure of
BLRs) was that the higher-ionization lines come from closer in to
the central source (Gaskell \& Sparke 1986). This has now been
widely confirmed by major observational campaigns studying a
number of objects.  Not only are the responsivity weighted radii
different, but the ``transfer functions'' (the light echoes seen
in response to a $\delta$-function in the photoionizing continuum)
of BLR~I and BLR~II have difference shapes (compare Krolik {\it et
al.} 1991 with Horne, Welsh, \& Peterson 1991).  While we now
recognize that BLR~I is ``stratified'' and seems to have a
different structure from BLR~II, it is important to note that the
distances of the different emitting regions from the black hole
are not very different.

\section{Spectropolarimetric Results}

Spectropolarimetry is proving to be a powerful tool for probing
the structure of AGNs.  Three results stand out in particular:

\begin{enumerate}
\item The percentage polarization can be different for the BLR and
the continuum.
\item The position angle (PA) of the polarization can also be
different for the BLR and the continuum.
\item Both the percentage polarization and the PA can vary {\it
across} the BLR line profiles (see, for example, Martel 1998).
\end{enumerate}

The first two results imply that the size of the scattering region
is comparable to the size of the BLR (and perhaps the scatterer is
mixed in with the BLR?).  The third result is very important for
BLR kinematics because it implies that {\it there is organized
bulk motion in the BLR}.

\section{How is the BLR Gas Moving?}

There is no consensus as to what the BLR is doing. Almost every
kind of model of BLR kinematics is still under active
consideration: random motions, Keplerian discs, infall, and
various sorts of outflows or winds.  I believe that understanding
BLR kinematics is crucial for understanding the role of the BLR in
AGNs and why NLS1s are different. My confident pre-1987
expectation was that BLR~I at least was moving radially outwards.
This was because of
\begin{itemize}
\item The blueshift of BLR I.
\item The existence of broad absorption line quasars.  Gas
causing blueshifted absorption lines must be moving away from the
quasar.
\end{itemize}
If the BLR is moving radially outwards then as the emission lines
vary in response to changes in the ionizing continuum we expect
the blue wings of the line to vary first, since this gas would be
closest to our line of sight.  I was therefore quite surprised to
find (Gaskell 1988$b$) that the wings of C~IV vary almost together
and the red wing leads the blue wing {\it slightly}. This has now
been confirmed for many objects (see, for example, Koratkar \&
Gaskell 1991).  There are several possible solutions to the
dilemma these results present:

1. {\it The Shifts Are Not Caused by Bulk Radial Motion of BLR
Clouds}

Electron scattering can cause a blueshift of line profiles
(Edmonds 1950).  This has been modeled more recently by Kallman \&
Krolik (1986) and Ferrara \& Pietrini (1993).  At least a modest
optical depth to electron scattering is needed ($\tau_{es} \sim
0.5)$.  This mechanism has the advantage that we {\it know} the
electrons are there because we see them in polarized light as the
so-called ``mirrors'' in some Seyfert 2s. Continuity requires that
the density of electrons increase inwards from the observed
``mirror'' to the BLR as $r^{-2}$.  The stellar wind model of
Taylor (1998; see also these proceedings) offers another possible
explanation of the blueshifting.

2. {\it There is Bulk Radial Outflow, but it is Not Causing the
Blue Wing to Vary First}

There have been a couple of proposals in this category.  These
include the disc/wind model of Chiang \& Murray (1996) and the
hydromagnetic outflow model of Bottorff {\it et al.} (1999).

\section{Line Emission from Discs}

If the dominant motion of the BLR is in a plane, the line width
depends on the orientation of the plane of motion ({\it e.g.}, the
disc) to the observer's line of sight. A face-on viewing
position has been a widely discussed explanation of the narrowness
of the BLR lines in NLS1s. Wills \& Browne (1986) showed that, for
radio-loud AGNs at least, there is indeed a strong correlation
between the FWHM of H$\beta$ and orientation.

If we are not viewing the disc face on, the Keplerian rotation
will make us see double-humped profiles. Such displaced humps have
long been seen, especially in the so-called `3C~390.3 objects,' but
the disc explanation of this line profile structure has been
controversial (see Gaskell \& Snedden 1999 and Sulentic {\it et
al.} 1999). Problems arise because the disc model makes some
definite predictions that were not verified:
\begin{itemize}
\item The blueshifted hump should always be stronger than the
redshifted hump.  In fact it has long been known that there are
cases where the {\it redshifted} hump is much stronger than the
blueshifted hump ({\it e.g.}, Osterbrock \& Cohen 1979) and our
statistical study of line profiles (Gaskell \& Snedden 1999)
shows that red peaks are about equally likely to be stronger than
blue peaks.
\item The BLR responds to changes in the photoionizing continuum.
If the source of ionizing radiation is located on the axis of
symmetry, the observer sees the red and the blue sides of the line
going up and down in intensity together.  Gaskell (1988$a$)
pointed out that this was not the case.
\end{itemize}
Despite my earlier objections I have now become convinced that we
{\it are} seeing the signature of disc emission in some BLR line
profiles. The {\it International AGN Watch} (IAW) monitoring of
3C~390.3 (Dietrich {\it et al.} 1998) showed that on a {\it
light-crossing timescale} both the red and blue sides of the
Balmer lines vary up and down together as predicted by the disc
model. The profile changes reported earlier were on a much longer
timescale and are not related to ionizing continuum changes.
Interestingly, profile changes in typical (non-NLS1) radio-quiet
AGNs are also independent of what the ionizing continuum is doing
(see Peterson, Pogge, \& Wanders 1999).

The line-profile statistics and long-term profile variability
observations can all be explained if there is emission from a
disc that is not azimuthally symmetric.  The asymmetries might
be hot spots or spiral wave patterns on the disc (see Gilbert
{\it et al.} 1999 for some possible models). In at least one
object (3C~390.3), a drift in wavelength consistent with orbital
motion has been seen (Gaskell 1996; Eracleous {\it et al.} 1997).

It might be said that NLS1s are very different from broad-line
radio galaxies, such as 3C~390.3, but it is important to recognize
that some objects displaying disc-like emission line profiles are
radio-quiet Seyfert galaxies.  Also, difference spectra of
``typical'' Seyferts often show the same disc-like signatures seen
in 3C~390.3 objects.  We have carried out a study of the
statistics of apparent structure in Balmer line profiles for
samples of radio-loud and radio-quiet objects (Gaskell \& Snedden
1999), and the results are consistent with the hypothesis that a
disc-like component of BLR emission is present in {\it all} AGNs,
regardless of type. Disc-like profiles are harder to detect in
narrower-line objects whether they be radio-loud or radio-quiet.
We believe the disc component is harder to detect in radio-quiet
objects simply because they tend to have narrower line profiles.

\section{Putting it All Together}

There are many results and complications that space does not
permit me to cover, but, while the jury is still out on many key
issues (notably the kinematics), I think a picture is emerging:
the BLR consists of two main components. One component (BLR~II)
arises predominantly in a very optically thick disc and the other
(BLR~I) arises predominantly from a more spherical distribution of
more optically thin gas. If this is correct the leading model is
therefore one of the disc-plus-atmosphere models proposed by
Shields (1977), Collin-Souffrin {\it et al.} (1980), and others.

At a conference like this we are focusing in on differences
between objects but it is important to realize the great {\it
similarities} between between members of the quasar family,
especially in continuum and BLR properties.  The differences are
mostly subtle, and it took a lot of work by many people over many
years to discover them.  There is also a continuum in the
distribution of differences.

Why do BLR properties change as the BLR~II lines get narrower?
({\it i.e.}, as we go to NLS1s)?  The thermalization of H$\beta$
and the increase in Si~III]/C~III] tells us that the density must
be higher in NLS1s. Since the FWHM difference is most pronounced
for the BLR~II lines, and variations in the FWHM of BLR~II are
demonstrably influenced by orientation for non-NLS1s (Wills \&
Browne 1986), NLS1s are presumably seen face-on. I offer the
following as a tentative attempt at an integrated picture of the
NLS1 phenomenon:
\begin{enumerate}
\item The driving factor is a denser circumnuclear environment.
\item This denser environment provides an enhanced black hole fueling
rate, as has been argued elsewhere in this conference.
\item The denser environment produces more obscuration of the central
engine and produces a narrower opening angle through which we see
BLR~II more face-on than in non-NLS1s.
\end{enumerate}

%\section{Acknowledgements}

\begin{ack}
I would like to thank all the conference organizers and especially
Thomas Boller for all their efforts in putting on this
particularly well focused workshop.
\end{ack}

%This is an example how to include a postscript file:

%\begin{figure}[htb]
%\centerline{\psfig{figure=mjf_chi2.ps,height=3.9truein,width=3.5truein,angle=0}}
%\caption{Caption text...}
%\end{figure}

% -------------------------------------------------------------------------

% -------------------------------------------------------------------------

\end{document}